\documentclass[preprint,aps]{revtex4}
\usepackage{graphicx}

\input{epsf}
\begin{document}
%\draft
\title{\bf The Raychaudhuri equations: a brief review }
\author{Sayan Kar \thanks{Electronic Address :
sayan@phy.iitkgp.ernet.in//sayan@cts.iitkgp.ernet.in}} 
\address{Department of Physics and Centre for Theoretical Studies,\\
Indian Institute of Technology, Kharagpur 721 302, INDIA}
\author{Soumitra SenGupta \thanks{Electronic Address :
tpssg@mahendra.iacs.res.in}} 
\address{Department of Theoretical Physics,\\
Indian Association for the Cultivation of Science, Jadavpur, Kolkata  
700 032, INDIA}
%\twocolumn[1
%\maketitle
%\widetext
%\parshape=1 0.75in 5.5in

\begin{abstract}
We present a brief review on the Raychaudhuri equations. Beginning
with a summary of the essential features of the original article
by Raychaudhuri and subsequent work of numerous authors, we move
on to a discussion of the equations in the context of 
alternate non--Riemannian spacetimes as well as other theories of gravity,
with a special mention on the equations in spacetimes with torsion 
(Einstein--Cartan--Sciama--Kibble theory). Finally, we give an overview of 
some recent applications of these equations in General Relativity, 
Quantum Field Theory, String Theory and the theory of relativisitic membranes. 
We conclude with a summary and provide our own perspectives on directions of 
future research.
\end{abstract}

%PACS number(s) :
%\pacs{}
%]

\maketitle
%\narrowtext 
\newpage
\vspace{.2in}

\section{The beginnings}
About half a century ago, General Relativity (GR) was young 
(just forty years old!), and even the understanding of the simplest solution, 
the Schwarzschild, was incomplete. Cosmology was virtually in its infancy, 
despite the
fact that the Friedmann--Lemaitre--Robertson--Walker (FLRW) 
solutions had been around for quite a while. The question
about the then--known exact solutions of GR, which worried the serious 
relativist quite a bit, concerned their singular nature. Both the Schwarzschild
and the cosmological solutions were singular. It is well--known
that the creator of GR, Einstein himself, was quite worried about
the appearance of singularities in his theory. Was there a way out? 
Was it correct to believe in a theory which had singular solutions?
Were singularities inevitable in GR? 

It was during these days in the early 1950's, Raychaudhuri
began examining some of these questions in GR. One of his
early works during this era involved the construction of a
non--static solution of the Einstein equations for a cluster of
radially moving particles in an otherwise empty space
{\cite{akr1953}}. A year before, he had also written an article
related to {\em condensations} in an expanding universe \cite{akr1953a} 
where he dealt with
cosmological perturbations  (in a sense, this article
deals with what is today known as {\em structure formation}). 
Subsequent to these papers, in 1955, appeared
{\em Relativistic Cosmology I}
{\cite{akr1955}}, which contains the derivation of the now--famous
{\em Raychaudhuri equation}. 

Fifty years hence, the Raychaudhuri equations have been discussed and 
analysed in a variety of contexts. Their rise to prominence
was largely due to their use (through the notion of geodesic focusing) 
in the proofs of the
seminal Hawking--Penrose singularity theorems of GR. Today, the importance of 
this set of equations, as well as their applicability in diverse
scenarios, is a well-known fact.
   
This article is a brief review on these equations. We shall 
deal with some selected aspects in greater detail. We, of course,
would like to emphasize that there are many topics which we
leave untouched or, barely touched. We hope to do justice to these
in a later, and more extensive article. 

The overall plan of this article is as follows. In the remainder of this
section, we shall recall the basic ideas and results in the 1955
paper and the singularity theorems. The next section introduces 
(with illustrative examples) the kinematical quantities 
(expansion, rotation and shear)
which govern the characteristics of geodesic flows and also outlines
the derivation and consequences of 
the equations. Section 3 considers the equations in 
alternative Riemannian and non--Riemannian theories of gravity ($R+\beta R^2$ theory and the 
Einstein--Cartan--Sciama--Kibble (ECSK) theory, in particular).
In Section 4, we give a glimpse of the diverse uses of these equations in 
contexts within, as well as outside the realm of GR. Finally, we present
a summary and provide our perspectives on possible future work. 

\subsection{The original 1955 paper}

The derivation of the Raychaudhuri equation, as presented in the 1955
article, is somewhat different from the way it is arrived at in
standard textbooks today. It must however be mentioned, that 
in a subsequent paper in 1957 {\cite{akr1957}}, Raychaudhuri presented further
results which bear a similarity with the modern approach to
the derivation. 
Let us now briefly summarise the main points of the 
original derivation of Raychaudhuri.

\noindent (a) Raychaudhuri's motivation behind this article is
almost entirely restricted to cosmology. He assumes the
fact that the universe is represented by a time--dependent
geometry but does {\em not} assume homogeneity or isotropy
at the outset. In fact, one of his aims is to see whether non--zero
rotation (spin), anisotropy (shear) and/or a cosmological constant 
can succeed in avoiding the intial singularity.   

\noindent (b) The entire analysis is carried out in the comoving frame
(in the context of cosmological line elements) 
--the frame in which the observer is at rest in the fluid.

\noindent (c) The quantity $R^4_4$ (spacetime coordinates in the
1955 paper are
labeled as $x^1,x^2,x^3,x^4$ with the fourth one being time),
is evaluated in two ways--once using the Einstein equations
(with a cosmological constant $\Lambda$) and, again, using the
geometric definition of $R^4_4$ in terms of the metric and
its derivatives. In the second way of writing this quantity,
Raychaudhuri introduces the definitions of shear and rotation.

\noindent (d) Finally, equating the two ways of writing $R^4_4$
the equation for the evolution of the expansion rate is
obtained. Note that the definition of expansion given in this
paper refers to the special case of a cosmological metric.

\noindent (e) Apart from obtaining the equation, the article also arrives at
the focusing theorem (though it is not mentioned with this name)
and some additional results (the last section).   

In the same year (1955), Heckmann and Schucking {\cite{hs1955}}, 
while dealing with
Newtonian cosmology arrived at a set of equations, one of which is
the Raychaudhuri equation (in the Newtonian case). Prompted
by this work, Raychaudhuri re--derived his equations
in a somewhat different way in an article where he also showed that
Heckmann and Schucking's work for the Newtonian case could be
generalised without any problems to the fully relativisitic
scenario. It must also be noted that Komar {\cite{komar1956}}, a year after
Raychaudhuri's article appeared, obtained conclusions similar to
what is presented in Raychaudhuri's article. Raychaudhuri
pointed this out in a letter published in 1957 {\cite{akr1957b}}.

Subsequently, in 1961, Jordan, Ehlers, Kundt and Sachs
wrote an extensive article on the relativistic mechanics
of continuous media where the derivation of the
evolution equations of shear and rotation  
seem to appear for the first time {\cite{ehlers}}.
Furthermore, for null geodesic flows, the kinematical
quantities: expansion, rotation and shear (related to the so--called
optical scalars) and the corresponding Raychaudhuri equations, 
were first introduced by Sachs {\cite{sachs}}. 

The Raychaudhuri equation is sometimes referred to as the
Landau--Raychaudhuri equation. 
It may be worthwhile to point out precisely, the work of
Landau, in relation to this equation.
Landau's contribution appears in his 
treatise {\em The Classical Theory of Fields} {\cite{ll}} and
is also discussed in detail in {\cite{lk1963,komar1956}}.
Working in the synchronous (comoving) reference frame, Landau
defines a quantity $\chi_{\alpha\beta}=\frac{\partial\gamma_{\alpha\beta}}{\partial t}$, where $\gamma_{\alpha\beta}$ is the 3--metric. Subsequently,
using the fact that $\chi^{\alpha}_{\alpha}=\frac{\partial}{\partial t}\gamma$,
where $\gamma$ is the determinant of the 3--metric, he writes down an
expression for $R_{0}^{0}$ and then, an inequality $\frac{\partial}{\partial t}
\chi^{\alpha}_{\alpha} + \frac{1}{6}\left (\chi^{\alpha}_{\alpha}\right )^2
\geq 0$. While deriving the inequality, Landau implicitly 
assumes the Strong Energy
Condition (though it is not mentioned with this name). Then, using it, he 
is able to show that $\gamma$ must necessarily go to zero within
a finite time. However, he mentions quite clearly that this does not
imply the existence of a physical singularity in the sense of curvature.
Though Landau's work captures the essence of focusing, he does
not explicitly mention {\em geodesic focusing}--moreover, he does not
introduce shear and rotation or write down the complete {\em equation}
for the expansion. 

Even though it was mentioned in {\cite{lk1963}} and {\cite{ehlers}},
Raychaudhuri's contribution found its true recognition only after the seminal 
work of Hawking and Penrose which appeared a decade later.
It was at that time, along with the proofs of the singularity theorems, the 
term {\em Raychaudhuri equation} came into existence in the physics
literature.  

\subsection{The singularity theorems}

It must be mentioned that Raychaudhuri did point out the connection of his 
equations to the existence of singularities in his 1955 article.
However, more general
results (based on global techniques in Lorentzian spacetimes) 
appeared in the form of singularity theorems following 
Penrose's work {\cite{rp1965}} and, then, Hawking's contributions 
{\cite{swh1965,swh1966}}. 
%It that the equations Raychaudhuri derived, could stand alone
%as {\em geometric statements} without any reference to the Einstein
%field equations. 
The crucial element of the singularity theorems is that
the existence of singularities is proved by using a minimal set of
assumptions (loosely speaking, these are : Lorentz signature metrics and
causality, the generic condition on the Riemann tensor components, 
the existence of trapped surfaces and energy conditions on matter). 
In fact, a precise definition of
`what is a singularity?' first appeared in the works of Hawking and
Penrose. The notion of geodesic incompleteness and its relation to
singularities (not necessarily curvature singularities) was also
born in their work. One should also realise that
the focusing of geodesics arrived at by Raychaudhuri and discussed in
much detail in later articles by other authors could be completely
benign (irrespective of any actual singularity being present in the
manifold). Thus, a singularity would always imply focusing of geodesics
but focusing alone cannot imply a singularity (also pointed out by Landau
{\cite{ll}}). We refrain from discussing the singularity theorems any
further here--excellent discussion on global aspects in gravitation as well as 
the Hawking--Penrose theorems are available in {\cite{wald,he1973,joshi}}.

\section{The geometry and physics of the equations}

Let us now review the basic ingredients and the derivation of the equations. 
First, of course, we need to know-- what do these 
equations deal with? In
a sentence, one may say that they are concerned about the {\em kinematics
of flows}. Flows are generated by a vector field--they are the
integral curves of the given vector field. These curves may be
geodesic or non--geodesic, though the former is more useful in
the context of gravity. Thus, a flow is a congruence of such curves
--each curve may be timelike or null or, in the Euclidean case,
have tangent vectors with a positive definite norm. One does not,
in the context of these equations, ask, how the flow is generated.
In other words, we are more interested, in deriving the 
kinematic characteristics of such flows. The 
evolution equations (along the flow) of the quantitites that characterise 
the flow in a given background
spacetime, are the Raychaudhuri equations. Historically speaking, it is
the equation for one of the quantitites (the expansion), which is termed
as the Raychaudhuri equation. However, in this article, 
we will refer to the full set of equations as Raychaudhuri equations.  

\subsection{Expansion, rotation, shear}

What quantities characterise a flow? If $\lambda$ denotes
the parameter labeling points on the curves in the flow, then, in order
to characterise the flow, we must have different functions of $\lambda$. 
In other
words, the gradient of the velocity field being a second rank tensor
is split into three parts : the symmetric traceless part, the antisymmetric
part and the trace. These define for us the shear, rotation and the expansion
of the flow. Specifically,  

\begin{equation}
\nabla_b v_a= \sigma_{ab} +\omega_{ab} +\frac{1}{n-1}h_{ab}\Theta
\end{equation}
where the symmetric, traceless part, the shear, is defined as $\sigma_{ab} =
\frac{1}{2} \left (\nabla_b v_a +\nabla_a v_b\right ) -\frac{1}{n-1} h_{ab}
\Theta$,
the trace, expansion, is $\Theta = \nabla_a v^a$ and the antisymmetric rotation 
is given as, $\omega_{ab}=\frac{1}{2} \left ( \nabla_b v_a - \nabla_a v_b\right ) $. n is the dimension of spacetime,
and $h_{ab} = g_{ab} \pm v_a v_b$ is the projection tensor (the plus
sign is for timelike curves whereas the minus one is for spacelike
ones). Also, correspondingly, $v_a v^a =\mp 1$. We shall discuss the
case of null geodesic congruences briefly, later. 

The geometric meaning of these quantities is shown through Figure 1 and
Figure 2. The expansion, rotation and shear are
related to the geometry of the cross sectional area (enclosing a fixed
number of geodesics)  orthogonal to the flow
lines (Figure 1). As one moves from one point to another,
along the flow, the shape of this area changes. It still includes the same
set of geodesics in the bundle but may be isotropically smaller (or larger), 
sheared or twisted. The analogy with elastic deformations or fluid
flow is, usually, a good visual aid for understanding the change in the 
geometry of this area. A recent, nice
discussion is available in \cite{poisson}. Earlier references where
these quantities are explained in quite some detail are {\cite{ellis1},\cite{ciufo}.  

\begin{figure}
\centerline{\epsfxsize=2.5in\epsffile{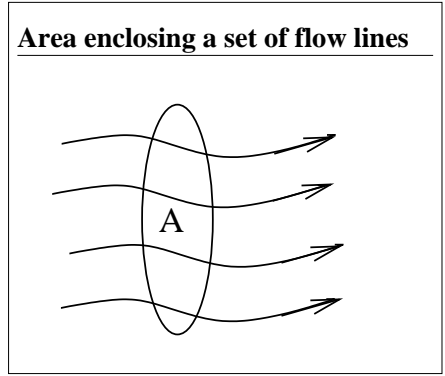}}
\caption{The cross sectional area enclosing a congruence of geodesics}
\end{figure}

\vspace{.1in}

\begin{figure}
\centerline{\epsfxsize=4.8in\epsffile{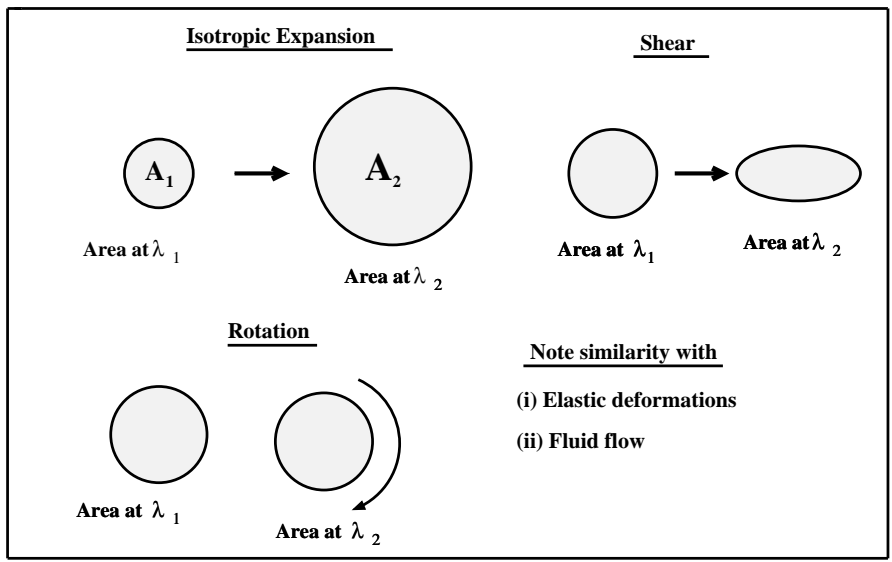}}
\caption{Illustrating expansion, rotation and shear}
\end{figure}

\subsection{Examples}

It is useful to illustrate these quantities with a set of
examples. We first choose to work with Schwarzschild spacetime. Our
examples here will involve 
(i) rotation--free timelike geodesic flows and (ii) timelike 
flows with all three kinematical quantities non--zero. We focus
on examples with non--zero shear and rotation because these
are not usually available in standard texts on GR. Our choice of examples are
primarily based on the problems suggested in a recent monograph by 
Poisson {\cite{poisson}}. In a third example (iii) we discuss
briefly a timelike geodesic flow in the FLRW universe. Finally, in (iv)
we briefly deal with geodesic flows in wormhole spacetimes.  
 
\noindent (i) From the Frobenius theorem \cite{poisson}, we know that 
hypersurface
orthogonal vector fields must necessarily have zero rotation though
the shear can have non--zero components.
In Schwarzschild spacetime, we construct a congruence which has the
above properties (i.e. it is irrotational but has nonzero shear and
expansion). Consider the vector field:

\begin{equation}
u^a \partial_a = \frac{1}{1-\frac{2M}{r}}\partial_t \pm \sqrt{\frac{2M}{r}}
\partial_r
\end{equation}
The geodesics corresponding to the above vector field are marginally
bound (i.e. $u_t=-1$) and the upper and lower signs refer to 
outgoing and incoming geodesics. It is easy to show that the
above vector field can be written as $u_a = \partial_a \phi$, where
$\phi (x^a) = constant$ would represent the hypersurface with respect to
which $u^a$ is orthogonal.  In fact, $\phi$ is the same as the new time
T (the proper time as measured by a freely falling observer starting
from rest at infinity and moving radially inward) 
used in the Painleve--Gullstrand representation \cite{poisson} 
of the Schwarzschild
line element.  

It is easy to calculate the expansion, which turns out to be:

\begin{equation}
\Theta = \pm \frac{3}{2} \sqrt{\frac{2M}{r^3}}
\end{equation}
Notice that the expansion is positive for outgoing and negative for incoming 
geodesics.
We can also find the nonzero shear tensor components
which are given as:

\begin{eqnarray}
\sigma_{tt} =\mp \frac{2M}{r^2} \sqrt{\frac{2M}{r}}\hspace{.1in};\hspace{.1in}
\sigma_{rr}= \mp \frac{\sqrt{\frac{2M}{r^3}}}{\left (1-\frac{2M}{r} \right ) ^2}\nonumber \\
\sigma_{tr} = \frac{\frac{2M}{r^2}}{1-\frac{2M}{r}} \hspace{.1in};\hspace{.1in}
\sigma_{\theta\theta}= \pm \sqrt{\frac{Mr}{2}} = \frac{1}{\sin^2 \theta}\sigma_{\phi\phi} \\
\end{eqnarray}
One can then check that the Raychaudhuri equations (given in the next sub--section) 
hold with the above expressions for the shear and expansion. 
 
\noindent (ii) We now move on to an example where all three kinematical
quantities are non--zero.
Consider the following vector field in Schwarzschild spacetime:

\begin{equation}
u^a \partial_a \equiv \frac{1}{\sqrt{1-\frac{3M}{r}}}\left (\partial_t +
\sqrt{\frac{M}{r^3}} \partial_{\theta} \right )
\end{equation}
where $M$ is the usual mass. It can be verified that the
geodesics corresponding to this vector field are timelike and they
are circular (r is a constant). We intend to calculate the expansion,
rotation and shear of the above vector field. 
The expansion is given as :

\begin{equation}
\Theta = \cot \theta \sqrt{\frac{M/r^3}{1-\frac{3M}{r}}}
\end{equation}
Notice that the expansion is positive in the northern hemisphere and 
negative in the southern hemisphere. 

Following the definition, one can show that
the rotation tensor for this vector field is given as:

\begin{equation}
\omega_{tr} = \frac{M}{4r^2} \frac{1-\frac{6M}{r}}{\left (1-\frac{3M}{r}\right )^{\frac{3}{2}}} = \sqrt{\frac{M}{r^3}} \omega_{r\theta}
\end{equation}
and the shear tensor is:

\begin{eqnarray}
\sigma_{tt} = \frac{M}{r^3} \sigma_{\theta\theta} = -\frac{M}{2r^3\sin^{2}\theta} \frac{\left (1-\frac{2M}{r}\right )}{\left (1-\frac{3M}{r}\right )} \sigma_{\phi\phi} \\ \nonumber
=-\sqrt{\frac{M}{r^3}} \sigma_{t\theta} = \frac{M}{r} \frac{\left (1-\frac{2M}{r}\right )^2}{\left (1- \frac{3M}{r}\right )} \sigma_{rr}= -\frac{1}{3} \cot \theta \sqrt{\frac{M^3}{r^5}} \frac{(1-\frac{2M}{r})}
{\left (1 -\frac{3M}{r} \right )^{\frac{3}{2}}} \\ \nonumber
\sigma_{tr} = \sqrt{\frac{M}{r^3}} \sigma_{r\theta} = -\frac{3M}{4r^2}
\frac{(1-\frac{2M}{r} )}{\left (1-\frac{3M}{r} \right )^{\frac{3}{2}}} 
\end{eqnarray}
One can further verify that here too, the Raychaudhuri equations (given below) 
are satisfied for the above quantities. One can also evaluate $\sigma^2-\omega^2$ and show that it is positive for $r>3M$. The expansion is defined for
this domain of r ($>3M$) and can diverge to negative infinity (focusing) at $\theta=\pi$ (south pole).

(iii) As a third example, we now quickly discuss the expansion, rotation and shear with respect to
the vector field $u^a\partial_a = \partial_t$ in the standard  
cosmological line element (FLRW). The shear and rotation are identically zero.
The expansion is given as: 

\begin{equation}
\Theta = 3\frac{\dot a}{a} = \frac{1}{\sqrt{a^6}}\frac{d}{dt} \left (
{\sqrt{a^6}} \right )
\end{equation}
Note that $a^6$ is the volume of the expanding 3--space. Hence the term
`volume' expansion is also used in the literature. 
Raychaudhuri, in his original article, defined the expansion in the this way. 
However, his treatment did include
nonzero shear and rotation because he did not assume, to start with,
maximally symmetric metrics on spatial slices representing $R^3$, $S^3$ or $H^3$. It may be
mentioned here that the equation for the expansion reduces to the
equation for $\frac{\ddot a}{a} = \frac{4\pi G}{3} \left ( \rho + 3 p\right )$. 

(iv) Our final example concerns the case of geodesic flows in a
traversable wormhole \cite{wormholes}. It is known that traversable wormholes
require energy--condition--violating matter. These are non--singular
spacetimes where the spatial slices resemble two asymptotically flat regions
connected by a throat. Thus a geodesic congruence passing through the
throat from one asymptotic region to the other would necessarily
tend to focus first, not quite reach a focal point, and then defocus.
For a typical wormhole, the line element is given as:

\begin{equation}
ds^2 = -\chi^2 (l) dt^2 + dl^2 +r^2 (l) \left (d\theta^2 +\sin^2 \theta d\phi^2
\right )
\end{equation}
where, for a wormhole, $\chi (l)$ is nonzero and finite for all $l$ and
$r(l=0)=b_0$ with $r (l\rightarrow \pm \infty ) \sim l$. 
It is easy to check that the expansion is proportional to $\frac{r'}{r}$ (the
prime denoting a derivative w.r.t. l). Thus, the expansion never becomes
negative infinity and thus there is no focusing. One can work out 
the expansion  for a typical example using $r(l) = \sqrt{b_0^2 + l^2}$
(the so--called Ellis wormhole) \cite{wormholes}. 

\subsection{The equations and the focusing theorem}

We now turn towards writing down the evolution equations for the
expansion, shear and rotation along the flow representing a 
timelike geodesic congruence. 
A fact worth mentioning here is that, these evolution equations (and their
generalisations) are essentially
{\em geometric statements} and are independent of any reference to the 
Einstein field equations. 

The modern (textbook) way to derive these 
equations (see \cite{wald}) is as follows. Consider the quantity
$v^c\nabla_c B_{ab}$ (where $B_{ab}= \nabla_b v_a$). Evaluate this
as an identity and then split it into its trace, antisymmetric and
symmetric traceless parts. The equations that emerge are the ones
given below (for n $\equiv$ the dimension of spacetime $=4$).   

\begin{equation}
\frac{d\Theta}{d\lambda} +\frac{1}{3}\Theta^2 +\sigma^2-\omega^2 =
-R_{ab}v^a v^b
\end{equation}

\begin{eqnarray}
\frac{d\sigma_{ab}}{d\lambda} = -\frac{2}{3}\Theta \sigma_{ab}
-\sigma_{ac}\sigma^{c}_{\,\,\,b} -\omega_{ac}\omega^{c}_{\,\,\,b}
+ \frac{1}{3}
h_{ab} \left (\sigma^2-\omega^2 \right ) \nonumber \\ + C_{cbad}v^c v^d +
\frac{1}{2} {\tilde R}_{ab} 
\end{eqnarray}

\begin{equation}
\frac{d\omega_{ab}}{d\lambda} = -\frac{2}{3}\Theta \omega_{ab} -
2\sigma^{c}_{\,\,\,[b}
\omega_{a]c} 
\end{equation}
where $\sigma^2=\sigma_{ab}\sigma^{ab}$, $\omega^2=\omega_{ab}\omega^{ab}$,
$C_{cbad}$ is the Weyl tensor and the quantity
${\tilde R}_{ab} = h_{ac}h_{bd}R^{cd} - \frac{1}{3}h_{ab}h_{cd}R^{cd}$. 

There are a few points to note here. Firstly, one must realise that
these are {\em not} equations but, essentially, identitites. Hence,
in some references {\cite{identity,zafiris,carter}} we find the usage 
{\em Raychaudhuri identity} or {\em Codazzi--Raychaudhuri identity} (in the
context of surface congruences to be discussed later) 
which is, indeed, rigorously correct. The identities, however become
equations once we use the Einstein equations or any other geometric
property (e.g. Einstein space, or vacuum, etc.) as an extra input.
However, we shall continue to use the term {\em equations} in this
article.

Furthermore, the equations are coupled, nonlinear and first order.
The equation for the expansion is of central interest (in the context
of the singularity theorems) and it
is rather straightforward to analyse. In mathematical parlance, it (the
equation for the expansion) is known as a {\em Riccati equation}.
Such equations can be transformed into a second order linear form
(more precisely a Hill--type equation or a harmonic oscillator equation
with a time-varying frequency) \cite{fjt1978,fjt1978a}. 
Redefining $\Theta = 3 \frac{F'}{F}$
one gets:

\begin{equation}
\frac{d^2 F}{d\lambda^2} + \frac{1}{3}
\left (R_{ab}v^{a}v^b +\sigma^2
-\omega^2 \right ) F =0
\end{equation}
The analysis of the expansion equation can be done using the above
form. One notes that the expansion $\Theta$ is nothing but the rate of
change of the cross--sectional area orthogonal to the bundle of geodesics. 
Therefore,
the expansion approaching negative infinity implies a convergence of the
bundle, whereas a
value of positive infinity would imply a complete divergence.
What are the conditions for convergence? Firstly, for convergence we
must have an initially negative expansion. Finally, with $F'$ negative 
we must end up at a zero of F (at a finite $\lambda$),
in order to have a negatively infinite expansion \cite{fjt1978,fjt1978a}.
Thus, the criterion for the existence of zeros in F at finite values 
of the affine parameter is what is required for convergence. 
Using the well-known Sturm comparison
theorems in the theory of differential equations one can
show that convergence occurs if :

\begin{equation}
R_{ab}v^{a}v^b +\sigma^2
-\omega^2  \geq 0
\end{equation}
Thus, rotation defies convergence, while shear assists it. The equation
for the evolution of the rotation $\omega_{ab}$, has a trivial solution
$\omega_{ab}=0$. The criterion for convergence then
becomes particularly simple for such hypersurface orthogonal congruences
(zero rotation) : $R_{ab} v^a v^b \geq 0$. 
This leads to {\em geodesic focusing}.

If we make use of the Einstein field equations and rewrite the
Ricci tensor in terms of the energy--momentum tensor $R_{ab} = T_{ab} -
\frac{1}{2} g_{ab} T$ the the so--called timelike convergence condition
becomes a condition on matter stress energy. This, given as, 
$\left ( T_{ab} - \frac{1}{2} g_{ab} T\right ) v^a v^b \geq 0$ is
known as the {\em Strong Energy Condition} (SEC). For a diagonal
$T_{ab}$ (with $T_{00} = \rho, T_{aa} = p_a$)  we must have $
\rho +p_a \geq 0, \rho + \sum_a p_a \geq 0 $ if the SEC is to be obeyed.
In other words, geodesic focusing encodes the simple statement that
if matter is attractive, geodesics must be eventually drawn towards
each other. This seemingly trivial statement is {\em proved} via
the focusing theorem. 

In the late seventies, Tipler {\cite{fjt1978,fjt1978a}} realised that the 
assumption of the SEC imposed to prove focusing and hence the existence of 
singularities could be further weakened. Among other results, he was able to 
show how, in the proof of the Hawking--Penrose theorem one might replace SEC
by the Weak Energy Condition (WEC: $T_{ab} v^a v^b \geq 0$  $\forall$ 
non--spacelike $v^{a}$). Tipler also introduced in his article, for the
first time, the notion of an averaged energy condition (the Averaged
Strong and Weak Energy Conditions (ASEC and AWEC) which are global
in nature. For instance the AWEC is obtained by integrating the
WEC along a non--spacelike geodesic and gives a number ($\int_{\lambda_1}^{\lambda_2} T_{ab} v^a v^b d\lambda \geq 0$). The question of whether a violation
of the energy conditions could lead to a non--singular solution was also
addressed by him. One should mention here that a few years before Tipler's
work, Bekenstein 
\cite{jdb1976}
and Murphy \cite{glm1973} had proposed certain non--singular spacetimes.
These models turned out to be singular, following the above--mentioned 
generalisations of the singularity theorems.
Tipler also quotes, in his paper, the well--known Epstein--Glaser--Yaffe
theorem in quantum field theory where it is said that there can exist a 
quantum state w.r.t. which the expectation value of the stress--energy 
tensor can be negative \cite{egy}. However, he does mention that the existence of 
{\em one} state w.r.t. which the $\langle T_{00}\rangle$ is less than zero 
cannot really lead to the prevention of singularities. Of late, however, 
energy--condition violations have been discussed in great detail in the 
context of wormholes {\cite{wormholes}, dark energy \cite{darkenergy}, 
braneworld models \cite{rs} and semi--classical gravity (quantum
field theory in curved spacetimes) \cite{scg}. 
The experimentally observed Casimir
effect (the feeble attraction between parallel, conducting capacitor plates)
\cite{casimir} 
has been cited as an example of the existence of negative energy
density (though, truly speaking this effect is concerned with negative
pressures as opposed to actual negative energy densities). 

The quest for further weakening the criteria for geodesic focusing and
thereby making the singularity theorems stronger was continued later
through the work of Borde \cite{ab1987} and Roman \cite{tr1988}. Further
work on issues related to geodesic focusing, singularites, energy conditions 
and causality violations have been 
carried out in {\cite{fjt1977,ce1980,rb1983,tr1986,viss1992}}.

\subsection{Null geodesic congruences}

The Raychaudhuri equations for null geodesic congruences 
were first derived by Sachs {\cite{sachs}} in 1961. Let us briefly
recall the salient features of these equations. It must be
mentioned, however, that these equations are not very different (in
structure as well as consequences thereof) from the
equations for timelike congruences. 

The central issue in the case for null geodesic congruences is the
construction of the transverse parts of the deviation vector and the 
spacetime metric. Assuming an affine parametrisation in the sense
$dx^a=k^a d\lambda$ with $k^a k_a=0$ and $k^a \xi_a=0$ (with $\xi^a$ being
the deviation vector) we realise that we are in trouble because of the
above two orthogonality relations. Naively writing $h_{ab} = g_{ab} + k_a k_b$
will not work here ($k^a h_{ab} \neq 0$). The transverse metric is thus
constructed by introducing an auxiliary null vector $N^a$ with $k_a N^a =-1$.
(the choice of minus one is by convention, the essence is that the quantity
must be non--zero). If we choose $k_a = -\partial_a u$ ($u=t-x$) then we can
have $N_a = -\frac{1}{2}\partial_a v $ and hence $h_{ab} = g_{ab} +k_a N_b
+k_b N_a$. This satisfies $k^a h_{ab}=0$ and $N^a h_{ab}=0$. Note that
$h_{ab}$ now is entirely two dimensional. Keeping this transverse
metric in mind we can proceed in the same way as for the timelike case by 
constructing ${\hat B}_{ab} = \nabla_b k_a$. We quote below the equation for 
the expansion:

\begin{equation}
\frac{d{\hat \Theta}}{d\lambda} + \frac{1}{2} {\hat \Theta}^2 + {\hat \sigma}^2
-{\hat \omega}^2 = -R_{ab}k^a k^b
\end{equation}
where the hatted quantitites are the expansion, rotation and shear
for the null geodesic congruence. The focusing theorem for null geodesic
congruences follows in the same way as for timelike congruences, 
with the null convergence condition $R_{ab} k^a k^b \ge 0$ being the
requirement. Using Einstein equations one can obtain the so--called
Null Energy Condition $T_{ab} k^a k^b \ge 0$. Similar to the case for
timelike congruences, we have corresponding equations for the evolution
of shear and rotation for null geodesic congruences too. These are
available in {\cite{wald,poisson}}.

\subsection{The acceleration term and non--affine parametrisations}

The discussions above were exclusively for {\em geodesic} congruences. 
In the non--geodesic (i.e. timelike or null congruences) case, it is obvious 
that there will be differences.
We state below, how the equation for the expansion changes via the addition
of the so--called acceleleration term. The equation is now given as:

\begin{equation}
\frac{d\Theta}{d\lambda} +\frac{1}{3}\Theta^2 +\sigma^2-\omega^2 - \nabla_a \left (v^b \nabla_b v^a \right ) =
-R_{ab}v^a v^b
\end{equation}
where the fifth term on the R. H. S. is the acceleration term. Notice that 
this term is zero for geodesic congruences (zero acceleration). The average 
distance
between the world lines in the flow is changed due to the divergence of the 
acceleration. The non--geodesic character of the flow also affects the
equation for the shear.  

For a non-affine parametrisation of null geodesic congruences, with
$k^a\nabla_a k^b = \kappa k^b$ ($\kappa$, a constant, defined through
the above equation) the definition of the expansion changes : $\Theta =
\nabla_a k^a - \kappa$ and the Raychaudhuri equation for the expansion
takes the form:

\begin{equation}
\frac{d{\hat \Theta}}{d\lambda} - \kappa \Theta + \frac{1}{2} {\hat \Theta}^2 + {\hat \sigma}^2
-{\hat \omega}^2 = -R_{ab}k^a k^b
\end{equation}
The new feature here is the presence of the linear (in $\Theta$) term. 
The conclusions on geodesic
focusing however do not change, except for differences in the 
values of the expansion. 

\subsection{A theorem for non--rotating singularity--free universes : 
Raychaudhuri's later papers}

Over the entire period of more than forty years, Raychaudhuri did not
work much on the equations which bear his name today.  
Interestingly, he came back to have a look at them once again in the late 
nineties with a series of papers \cite{akr1998,akr2000}.  In these articles, 
he constructed a 
theorem on non--rotating singularity--free universes. The main inspiration
behind these papers were the singularity--free 
cosmological solutions due to Senovilla \cite{senovilla},
which created a lot of interest and curiosity among relativists in the
1990's. In fact, the work of Raychaudhuri sets out to
show that these solutions may not quite be physically relevant, though
surely, mathematically correct. Let us now briefly recall the theorem
of Raychaudhuri.  

The basic premise of \cite{akr1998,akr2000} is the use of {\em spacetime
averages} of quantitites, defined as:

\begin{equation}
\langle \chi \rangle = \left [ \frac{\int_{-x_0}^{x_0}\int_{-x_1}^{x_1}\int_{-x_2}^{x_2}\int_{-x_3}^{x_3} \chi \sqrt{\vert g \vert} d^4 x}{\int_{-x_0}^{x_0}\int_{-x_1}^{x_1}\int_{-x_2}^{x_2}\int_{-x_3}^{x_3}\sqrt{\vert g \vert} d^4 x}\right]_{lim x_{0,1,2,3}\rightarrow \infty}
\end{equation}
Raychaudhuri shows that for a singularity--free non--rotating universe, open in
all directions, the spacetime average of all stress--energy invariants, including the energy density vanishes. The proof is worked out using the 
spacetime averages of the scalars that appear in the Raychaudhuri equation
for the expansion. Following the statement of the theorem, Raychaudhuri
claimed that an observationally consistent universe cannot have a
zero average density and hence one must necessarily give up the hope of
having singularity--free solutions. Subsequent to Raychaudhuri's work,
Saa and Senovilla wrote a couple of comments which were mainly 
concerned with the
{\em converse} of Raychaudhuri's theorem and the question whether 
spacetime averages could really be a property which can distinguish between
singular and non--singular models. In fact, one must note that the
theorem does {\em not} say that if the spacetime averages are zero the 
spacetime must be non--singular. Therefore, the theorem cannot really
be used to distinguish between singular and non--singular cosmological
models. Furthermore, Senovilla, in his comment, made a {\em conjecture} 
that {\em spatial
} and not {\em spacetime} averages could be a distinguishing property
between singular and non--singular models. More precisely, his claim was
that for non--singular, non--rotating, globally hyperbolic and everywhere 
expanding models where SEC holds, the spatial averages of all stress--energy
invariants must vanish. Therefore, if such spatial averages do not vanish 
then the model must be singular. More details and a recent proof of the 
conjecture is available
in the article by Senovilla in this volume \cite{jose}. Despite the conflict between
whether {\em spatial} or {\em spacetime} averages was the crucial
distinguishing factor between singular and non--singular models, it goes 
without saying that Raychaudhuri's idea about {\em averages} being a deciding factor
will surely be remembered as a lasting contribution apart from his coveted
equations.

\section{The equations in different geometries and different theories of 
gravity} 

The Raychaudhuri equations discussed in the previous sections do not
change as long as we respect the Riemannian (pseudo--Riemannian) metric
structure of space or spacetime. However, in an alternative theory of
gravity, the Einstein field equations are of course different and hence
the relation between the energy--momentum tensor and other geometric
quantities do change. This results in a modification of the consequences
that arise while analysing these equations. On the other hand, if the
usual Riemannian structure of spacetime changes, such as in the case of
spaces with torsion (Einstein--Cartan--Sciama--Kibble theory), then 
the Raychaudhuri equations are surely different. We shall give an
example of both these scenarios in the following two subsections.
The equations in spacetimes with torsion is discussed in relatively
greater detail because, this is one scenario where we actually notice
a {\em generalisation} of the usual Raychaudhuri equations.    

\subsection{Metric theories with symmetric connections}
As mentioned above, for theories with a symmetric connection, the 
Raychaudhuri equations do not change, though the R.H.S. of the
expansion equation when written using matter stress-energy can be
very different from what it is in GR. 
This change
can surely affect geodesic focusing. It must be noted here that any change
in the conclusions about geodesic focusing in this case is inherently due
to the {\em new solutions} (spacetime geometries) of the modified 
Einstein equations for a given stress--energy.  The abovementioned 
aspects  have been
analysed in the context of 
the Brans--Dicke and Hoyle--Narlikar theories \cite{akr1974,sb1974,akr1975}  as
well as the $R+\beta  R^2$ 
theory \cite{kung1}, on which we focus our attention below.  

The study of the Raychaudhuri equations in the presence of higher curvature 
terms in the action  has been a subject of interest for a 
long time\cite{muller}.
The equation for the expansion, in a spacetime background solution 
of $R + \beta R^2$  gravity in the presence of matter ($\rho
\propto a(t)^{-n}$) has been analysed in \cite{kung}.  
The Strong Energy Condition (SEC) is examined and it is shown that
the condition for a big-bang singularity changes in such a scenario.
Recall that in GR, the singularity theorem 
indicates the inevitable presence of singularities when the 
SEC (as well as some other conditions) are satisfied. 
However, at the microscopic scale, in the evolution of the expansion of 
a geodesic congruence, quantum effects are expected to show up -- which, in 
turn, may completely change classical predictions.
One such quantum effect leads to the appearance of higher curvature terms, 
namely, quadratic gravity \cite{kung1}.  
Even though it might seem that the analysis in \cite{kung} for 
quadratic gravity is classical, it is, in a broader sense, 
a semi-classical analysis.
Classical solutions of quadratic gravity have been studied to explore the 
nature of singularities and it has been shown that 
the big-bang singularity may be
avoided \cite{whitt}.  In \cite{kung}, quadratic gravity, 
i.e. $R+\beta R^2$ gravity, was considered as a
a backreaction effect on pure Einstein's gravity . 
It was found that the SEC  for $R + \beta R^2$ gravity is
different from that of Einstein gravity. We briefly summarize the results of 
\cite{kung} below.

In GR, the SEC 
follows from the use of Einstein's equations through the relation:

\begin{equation} R_{ab}v^a v^b = 8\pi G\left[ T_{ab} -
{1\over 2} Tg_{ab}\right]v^a v^b  \geq 0
\end{equation}

for all timelike $v^a$.
This, as mentioned before, leads to focussing.  

The author in \cite{kung} is primarily interested in the cosmological 
singularity.  
To analyse the effects of the quadratic terms one only needs to find 
the new expression for the Ricci tensor 
$R_{ab}$ in terms of the energy momentum tensor 
$T_{ab}$. Therefore, we require the modified
field equations for quadratic gravity. It is known that this equation is 

\begin{equation} 
- G_{ab}+16\pi
G\beta\left({1\over 2}R^2g_{ab} - 2RR_{ab} 
-2R_{;n}^{;n}g_{ab} + 2R_{;a;b}\right) = -8\pi
GT_{ab}
\end{equation}
In a perturbative analysis, we are primarily interested in the first order 
contribution from $\beta R^2$
to the Raychaudhuri equation. After some algebra one arrives at,
\begin{eqnarray}
 R_{ab}v^a v^b = \left[ \tilde G(T_{ab} - {1\over 2}Tg_{ab}) \right.\nonumber \\
\left. 2\beta{\tilde G}^2 \left ( {1\over 2} T^2 g_{ab}{\tilde G} -2TT_{ab} 
{\tilde
G} - T_{;n}^{;n}g_{ab} - 2T_{;a;b}\right) \right ]
v^a v^b+ O(\beta^2 ) 
\end{eqnarray}
where $\tilde{G} = 8{\pi} G$.\\
Comparing the above equation with that in Einstein gravity, we find that an
{\em effective} energy momentum tensor appears in the RHS of the above 
equation. 
Assuming a matter dominated universe
with a characteristic dependence on the scale factor (i.e., $\rho =
{\rho_0\over a^n} $) and a local  conservation of $T_{ab}$ leads to $p
= {n-3\over 3}{\rho_0\over a^n}$.
Thus, finally, we have,

\begin{eqnarray}
R_{ab}v^a v^b= 
-{n-2\over 2}\tilde G\rho_n + \beta{\tilde G}^2\left[ 3n(n-1)(n-4)
\tilde G \rho_n^2  + 6 n^2(n-4)ka^{-2}\rho_n\right] + \nonumber \\
(A^2 -1) \left[ -{n\over 3}\tilde G\rho_n + \beta{\tilde G}^2\left[
2n^2(n-4) \tilde G \rho_n^2  + 4 n(n+2) (n-4)ka^{-2}\rho_n\right]
\right] 
\end{eqnarray}
The above expression is the modified term which appears in the R. H. S. of
the Raychaudhuri equation for the expansion. 
It is thereafter analysed in the context of different values of $n$
and the possiblities of avoiding the big--bang singularity or its inevitable 
occurence are pointed out for different cases.

We repeat once again that for {\em any} modified theory of gravity 
(eg. Brans--Dicke,
Einstein--Gauss--Bonnet in higher dimensions, induced gravity, low energy
effective stringy gravity etc. ) as long the Riemannian structure of spacetime
is respected, changes in conclusions related to geodesic focusing can arise only
through the modified field equations and its use in the convergence condition. 

\subsection{ The Raychaudhuri equation in spacetimes with torsion}
The symmetric nature of the affine connection is one of the 
underlying assumptions of Riemannian geometry. This fact is also assumed
while constructing GR.
An antisymmetric connection may originate from the presence of spin matter 
fields in spacetime leading to a
transition from the {\bf V}$_4$ to {\bf U}$_4$ manifolds \cite{schouten}.
This asymmetric part is known as torsion.
Generalization of the theory of gravity in such a spacetime with torsion was 
proposed by
Einstein--Cartan--Sciama--Kibble (ECSK).\cite{hehl}.
Absence of any experimental signature of torsion however is the primary 
criticism of such models
although there has been a renewed  theoretical interest in the context of
superstring theories \cite{GSW} where spacetime torsion appears in the form 
of a massless string mode.
 The third rank field strength of the massless second rank antisymmetric 
tensor field of string theory (known as the Kalb-Ramond field) is 
identified with spacetime torsion in the
low energy limit of the theory. Torsion also appears naturally in
a theory of gravity where twistors are used
\cite{HOW} as well as in the supergravity scenario where
torsion, curvature and matter fields are treated in an analogous way
\cite{LOS}.
It has been shown in several articles \cite{we} that in string inspired models 
such a 
background with torsion results in a
departure from the experimentally predicted values of the well known 
phenomena like gravitational lensing,
perihelion precession of planetary orbits, gravitational redshift, 
rotation of the plane of polarization of the
distant galactic radiowaves etc.
All the above arguments, and several more, compel us to include torsion
in any comprehensive theory of gravity.
In the context of geodesic congruences it has further been shown that the 
kinematical quantities-- shear, rotation, acceleration, expansion and their
evolution equations are modified by the presence of torsion.
This naturally leads to a generalization of the
Raychaudhuri equations in presence of torsion leading to a more general 
understanding of the phenomenon of geodesic focusing.
Here we give a brief review of the work presented in \cite{capoz} where the role of torsion in modifying 
the Raychaudhuri equations and some of its implications have been discussed.

The torsion tensor $T_{ab}^{\phantom{ab}c}$ which is the antisymmetric part
of the affine connection $\Gamma_{ab}^{c}$, is given by,

\begin{equation}
\label{t1}
T_{ab}^{\phantom{ab}c}=\frac{1}{2}\left(\Gamma_{ab}^{c}-\Gamma_{ba}^{c}\right)
\equiv\Gamma_{[ab]}^{c}\,,
\end{equation}
where $a,b,c = 0,\dots 3$.

In GR, $T_{ab}^{\phantom{ab}c}$ is postulated to be zero.

From the torsion tensor one constructs the contortion tensor as,

\begin{equation}
\label{t3}
K_{ab}^{\phantom{ab}c}=-T_{ab}^{\phantom{ab}c}-T^{c}_{\phantom{c}a b}
+
T^{\phantom{a}c}_{b\phantom{c}{a}}=-K^{\phantom{a}c}_{a\phantom{c}b}\,,
\end{equation}
This leads to a general expression for the affine
connection given as, 

\begin{equation} \label{t5}
\Gamma_{ab}^{c}=\left\{^{c}_{ab}\right\}-K_{ab}^{\phantom{ab}c}\,,
\end{equation}
where $\left\{^{c}_{ab}\right\}$ is the symmetric part of the connection
(Christoffel symbols). With this modified connection
the commutator of the covariant derivatives of a scalar field $\phi$
is
\begin{equation}\label{scalarder}
 \tilde \nabla_{[a}\tilde\nabla_{b]}\phi=-T_{ab}^{\phantom{ab}c}\tilde\nabla_{c}\phi;
\end{equation}
which is zero in absence of torsion.\\
Similarly for a vector $v^a$ the commutator of the derivative gives,
\begin{equation}\label{doppiader1}
 \tilde \nabla_{[a}\tilde\nabla_{b]} v^{c}=
  R_{abd}^{\phantom{abd}c}v^d
 -2T_{ab}^{\phantom{ab}d}\tilde\nabla_{d}v^c,
 \end{equation}
 where the  Riemann tensor is defined as,
\begin{equation}\label{riemann}
 R_{abc}^{\phantom{abc}d}=\partial_{a}\Gamma_{bc}^{d}-\partial_{b}\Gamma_{ac}^{d}
 +\Gamma_{ae}^{d}\Gamma_{bc}^{e} - \Gamma_{be}^{d}\Gamma_{ac}^{e}.
\end{equation}
The contribution of torsion, to the Riemann tensor, is explicitly given 
through the following expression:

\begin{equation}\label{riexpanded}
R_{abc}^{\phantom{abd}d} =R_{abc}^{\phantom{abd}d}(\{\}) -
 \nabla_{a}K_{bc}^{\phantom{b]c}d} +  \nabla_{b}K_{ac}^{\phantom{ac}d}
+ K_{ae}^{\phantom{ae}d}K_{bc}^{\phantom{bc}e}-
K_{be}^{\phantom{be}d}K_{ac}^{\phantom{ac}e}
\end{equation}
where $R_{abc}^{\phantom{abc}d}(\{\})$ is the tensor generated by
the Christoffel symbols. The symbols $\tilde\nabla$ and $\nabla$
have been used to indicate the covariant derivative with and
without torsion respectively.

From Eq.(\ref{riexpanded}), the expressions for the Ricci tensor
and the curvature scalar are
\begin{equation}\label{ricci}
 R_{ab}= R_{ab}(\{\}) - 2\nabla_{a}T_{c} + \nabla_{b}K_{ac}^{\phantom{ac}b}
+ K_{ae}^{\phantom{ae}b}K_{bc}^{\phantom{bc}e}-
2T_eK_{ac}^{\phantom{ac}e}
\end{equation}
and
\begin{equation}\label{curvscalar}
 R=R(\{\}) - 4 \nabla_{a}T^{a} + K_{ceb}K^{bce} - 4 T_aT^a.
\end{equation}
where,
\begin{equation}\label{vector2}
  T_a= T_{ab}^{\phantom{ab}b}.
\end{equation}
$T_a$ can be either time-like, space-like or light-like.

\subsubsection{Contributions of torsion to shear,
expansion, vorticity and acceleration}\label{svea}
Beginning with the
behaviour of fluids and moving on to the initial singularity problem in 
cosmological models,
the Raychaudhuri equations have been shown to be the key equation to explore 
the role of
torsion in such diverse phenomena.\cite{drls},\cite{trautman},\cite{stewart},\cite{esposito}.
In \cite{ddrpss} an inflationary  model of the universe in the context of ECSK theory
was considered.
\cite{Ellis} and \cite{palle} addressed the crucial role of Raychaudhuri 
equation in the context of  a gauge invariant formalism for
cosmological perturbations in theories with torsion. We now explain the
kinematical quantities mentioned before, in a scenario with torsion.\\
Torsion in a
space-time modifies the definition of kinematical quantities.
The covariant
derivative of the four velocity $U_a$ \cite{he1973} can be decomposed as,
\begin{equation}\label{4vel}
 \tilde\nabla_{a}U_{b}= \tilde\sigma_{ab} + \frac{1}{3} h_{ab}
 \tilde\Theta+\tilde\omega_{ab} -U_{a}\tilde a_{b}
\end{equation}
where $h_{ab}=g_{ab} +U_a U_b$ and
\begin{equation}\label{expansion}
 \tilde\Theta= \tilde\nabla_{a}U^{a}=\Theta - 2T^{c}U_{c},
\end{equation}
\begin{equation}\label{shear}
\tilde\sigma_{ab}=h_a^c h_b^d\tilde\nabla_{(c}U_{d)}=\sigma_{ab} +
2h_a^c h_b^dK_{(cd)}^{\phantom{(ab)}e}U_{e},
\end{equation}
\begin{equation}\label{vorticity}
 \tilde\omega_{ab}=h_a^c
h_b^d\tilde\nabla_{[c}U_{d]}=\omega_{ab} + 2h_a^c
h_b^dK_{[cd]}^{\phantom{[cd]}e}U_{e},
\end{equation}
and the acceleration
\begin{equation}\label{acceleration}
  \tilde a_{c}= U^{a}\tilde\nabla_{a}U_{c}= a_{c}+
  U^{a}K_{ac}^{\phantom{ac}d}U_d.
\end{equation}
The quantities without the tilde are the value of the corresponding expressions in a spacetime without
torsion.
\subsubsection{The Raychaudhuri
equation}
Using the identity for the four-velocity $U_a$ ($U_a U^a =-1$),
\begin{equation}\label{dd}
 U^{b}\tilde\nabla_{c}\tilde\nabla_{b}U_{a}=\tilde\nabla_{c}
 (U^{b}\tilde\nabla_{b}U_{a}) - \tilde\nabla_{c}U^{b}
 \tilde\nabla_{b}U_{a}
\end{equation}
and from Eq.(\ref{riemann})
\begin{equation}\label{ddd}
U^{b}\tilde\nabla_{c}\tilde\nabla_{b}U_{a}=
U^{b}\tilde\nabla_{b}\tilde\nabla_{c}U_{a} +
R_{cba}^{\phantom{cba}d}U_{d}U^{b} - 2
U^{b}T_{ab}^{\phantom{ab}c} \tilde\nabla_{d}U_{c}
\end{equation}
we find the equation

$$ \frac{1}{3}h_{ca}\tilde\Theta +
\tilde\sigma_{ca}+\tilde\omega_{ca}- U_{c}\tilde a_{a}=
\tilde\nabla_{c}\tilde a_{a} $$
$$-\Bigg( \frac{1}{9}h_{ca}\tilde \Theta
+\frac{2}{3}\tilde\Theta\tilde\sigma_{ca} +
\frac{2}{3}\tilde\Theta\tilde\omega_{ca} +
2\tilde\sigma_{c}^{b} \tilde\omega_{ba} $$
$$\tilde\sigma_{c}^{b} \tilde\sigma_{ba} +
\tilde\omega_{c}^{b} \tilde\omega_{ba} -
\frac{1}{3}U_{c}\tilde\Theta\tilde a_{a} - U_{c}\tilde
a^{b} \tilde\sigma_{ba}-U_{c}\tilde a^{b}
\tilde\omega_{ba}\Bigg) $$
\begin{equation}\label{penultima}
-R_{cba}^{\phantom{cba}d}U_{d}U^{d}-2 U^{b}T_{ab}^{\phantom{ab}c}
\left(\frac{1}{3}h_{dc}\tilde\Theta +
\tilde\sigma_{dc}+\tilde\omega_{dc}- U_{d}\tilde
a_{c}\right).
\end{equation}
Contracting the indices in Eq.(\ref{penultima}), one obtains
a general expression for the Raychaudhuri
equation for the expansion, in the presence of torsion as,
\begin{eqnarray}\label{raychau}
\dot{\tilde\Theta}=\tilde\nabla_{c}\tilde a^{c}
-\frac{1}{3}\Theta^{2} -\tilde\sigma^{ab}\tilde\sigma_{ab}
+ \tilde\omega^{ab}\tilde\omega_{ab} - R_{ab}U^{a}U^{b} - 2
U^{b}T_{ab}^{\phantom{ab}d}
 \left(\frac{1}{3}h_{d}^{a}\tilde\Theta \right . \nonumber \\ +
\left . \tilde\sigma_{d}^{a}+\tilde\omega_{d}^{a}- U_{d}\tilde
a^{a}\right)\, 
\end{eqnarray}
This is the most general form of Raychaudhuri equation in
presence of torsion. Simpler versions of this equation have
been discussed in \cite{stewart}, \cite{tafel2},
\cite{esposito},\cite{raych}. It is obvious that there would be
corresponding equations for shear and rotation too, which we do not
mention here.
Interestingly, it can be shown that if we have torsion as a phantom field
(a scalar field with a negative kinetic energy term) 
through the dual of the third 
rank tensor field strength of a 
string inspired Kalb--Ramond field, one can get a positive 
contribution in the R.H.S. of 
the above equation, which, in turn, may be useful in eliminating 
singularities.  

\section{The equations in diverse contexts}

\subsection{General Relativity and Relativistic Astrophysics}

We have already discussed one of the main uses of the Raychaudhuri
equations--the geodesic focusing theorem. Though it is an entirely
geometric result, its use is mostly confined to the domain of GR. 
It is also quite obvious
that the null version of these equations are useful in the
study of gravitational lensing. The focal point in this case is nothing
but the intersection of trajectories representing light rays and is known as 
the caustic of the bundle of trajectories. The optical scalars (Sachs 
scalars) are the quantitites of interest and their evaluation enables
us to understand the nature of the null geodesic flow (light ray bundles).
Is there anything else one can say apart from focusing, which will
be relevant for lensing. Some of these issues have been addressed in \cite{void2000}. For example, it is possible to rewrite the 
Raychaudhuri equation for
null geodesics in a form involving the angular diameter distance $d_A$:

\begin{equation}
\frac{d_{A}''}{d_A} = -\frac{1}{2}R_{00} -\frac{1}{2} \sigma^2
\end{equation}
where $\frac{d}{d\lambda} \ln d_A =\frac{1}{2}\Theta$. Thus, for the case
of zero shear one can determine $d_A$ entirely in terms of geometry.
In \cite{void2000} the authors have asked the question `how do voids affect
light propagation?' and made use of the above relation to provide an 
answer.

A second example within relativistic astrophysics, is a study of
crack formation \cite{herrera,prisco} using this equation. This is an interesting application, quite unique-- but not pursued much later. The basic question
here is --  when will a spherical object develop cracks? 
Appearance of total radial forces of different signs
in different regions in a perturbed configuration leads to the 
occurrence of such cracking, which, in turn leads to local anisotropy of
fluid/emission of incoherent radiation. Using the Einstein and Raychaudhuri 
equations it is possible to express the net radial force R (after 
perturbation)  in terms of $\frac{d\Theta}{d\lambda}$, $p_i$, $\rho$, $g_{ij}$.
In this way a criterion for cracking can be obtained.

In 1995, Jacobson \cite{jacobson1} in a rather unique paper demonstrated how 
one might view the Einstein field equation as a thermodynamic equation of state.
In this calculation, Jacobson starts out by arguing that energy flux across a
causal horizon is some kind of heat flow and the entropy of the system 
beyond is proportional to the area of the horizon. The heat flux $\delta Q$
is related to the stress--energy tensor whereas the area variation is related
to the expansion of a bundle of null geodesics. He then makes use of the 
Raychaudhuri equation in its linearised form (ignoring the $\Theta^2$ term) to
write down the expansion as an integral over the $R_{ij}k^ik^j$. Thereafter,
with the input of the entropy--area relation he arrives at
the Einstein equation as an equation of state. It is important to note here
that Jacobson uses the {\em geometric} content of the Raychaudhuri
equation and views it as more fundamental than the Einstein equation, which,
in his approach is a {\em derived} relation.
Extensions of Jacobson's work in the setting of non--equilibrium thermodynamics
have appeared recently \cite{jacobson2}. 

Scattered around the literature, are innumerable instances where the
equations have been used in the context of GR and astrophysics.
Prominent among them is its use in black hole physics--in studying 
the properties of black holes
and in deriving the laws of black hole mechanics (see \cite{poisson} and
references therein).  
Apart from black holes, we mention, {\em en passant}, a few other situations 
which we thought were interesting: 
(i) use in a fluid--flow description of density irregularities in cosmology \cite{lyth}, (ii) quantum 
gravitational optics and an effective Raychaudhuri equation \cite{nouri2006}, 
(iii) magnetic tension and gravitational collapse \cite{tsagas}, (iv) 
the effective Einstein and Raychaudhuri equations derived from 
higher dimensional warped braneworld models 
\cite{maartens}.

\subsection{The Capovilla--Guven equations for relativistic membranes}

An obvious question, which was never asked till the work of Capovilla
and Guven {\cite{cg1995}} appeared in the scene is: what happens if we consider
a congruence of extremal surfaces instead of geodesics (extremal curves)? 
Are there similar Raychaudhuri equations? 

To address this issue, we must first find out how to generalise the
notions of expansion, rotation and shear for the case of a family of
surfaces. Unlike a curve, a surface is parametrised by more than one
parameter. Thus, it is natural to imagine an expansion, a rotation and
a shear along each of these independent directions. Introducing a
separate label for the surface coordinates, we find that we now have
$\Theta^a$, $\Sigma^a_{ij}$ and $\Omega^{a}_{ij}$ (here $i,j...$ denotes
the indices representing the normals to the surface). The relation 
between the $i,j$  indices and the spacetime indices (say $\mu,\nu$)
is given through the embedding of the surface. Let us now make things
more concrete and explicit. 

Define a D--dimensional  surface in a N dimensional background 
through an embedding $x^{\mu} = x^{\mu} (\xi^a)$.
$E^{\mu}_{a}$ constitute the tangent vector
basis chosen such that $g_{\mu\nu}E^{\mu a}E^{\nu b} = \eta_{ab}$ (a,b run
from 1 to D). $n^{\mu i}$ are the normals, with $g_{\mu\nu}n^{\mu i}n^{\nu j}
=\delta^{ij}$ (i,j run from 1 to $N-D$). Also $g_{\mu\nu}n^{\mu i}E^{\nu a}=0$.
 $K^{ab i}$ are the $N-D$ extrinsic curvatures (one along each normal direction)
The embedded surface is minimal provided $Tr(K_i) =0$. 

With the above definitions, one can follow the derivation for the case
of geodesic curves and obtain the equation for the generalised expansion 
(assuming zero values for the $\Sigma_{ij}$ and $\Omega_{ij}$:

\begin{equation}
\nabla^a\Theta_a +\frac{1}{N-D}\Theta_a\Theta^a + \left (M^2\right )^i_i =0
\end{equation}
where

\begin{equation}
\left (M^2 \right )^i_i = K^{ab i}K_{ab i} + R_{\mu\nu\rho\alpha}
E^{\mu a}E^{\rho}_a n^{\nu i}n^{\alpha}_i
\end{equation}
Note that the above equation is a partial differential equation--the reason 
being that we require more than one parameter to describe a surface. 
Structurally, the equation is similar to the original
Raychaudhuri equation for the expansion -- one can easily note this by comparing the two.
However, there are crucial differences -- one of which involves the appearance of extrinsic curvature
terms. It is possible to rewrite this equation in a second order form by choosing $\Theta_a=\nabla_a F$. We obtain:

\begin{equation}
\nabla_a\nabla^a F + \left ( M^2 \right )^i_i F = 0
\end{equation}
which, resembles a variable--mass wave equation on a D--dimensional surface  with a non--trivial
induced line element. For special cases, one can work out focusing criteria, although the
notion of focusing will be largely different for surface congruences. For the case of string
world--sheets, it is possible to make use of the conformal character of any two--dimensional
line element and rewrite the equation as:

\begin{equation}
\frac{\partial^2 F}{\partial \sigma^2} - \frac{\partial^2 F}{\partial \tau^2}  + \Omega^2 (\tau,\sigma) \left ( M^2 \right )^i_i F = 0
\end{equation}
where $\tau,\sigma$ are the coordinates on the string world sheet and the induced metric is
$ds^2 = \Omega^2 \left (-d\tau^2 +d\sigma^2\right )$.

It can be shown {\cite{sayan1} that the criterion for focusing is given as
(where ${}^2 R$ is the Ricci scalar of the worldsheet) :

\begin{equation}
-{}^2 R +R_{\mu\nu}E^{\mu a}E^{\nu}_{a} >0
\end{equation}
which is a requirement for the function F to have zeros. However, there are many questions 
related to focusing of surfaces which remain un--answered. Some of these have been recently
addressed in {\cite{sayan2}}. Earlier references on examples of solutions of the Raychaudhuri
equations for surface congruences and related issues are available in  {\cite{zafiris,carter,sayan3}}.
  
It must be mentioned here that the above generalised equation for the expansion vector field
is a subset of a full set of equations which include those for the generalised shear and rotation
too. Moreover, the analysis is entirely for extremal membranes of Nambu--Goto type. Here too,
if the action changes, we find that the equations change--an example is available in \cite{sayan4}.

\subsection{Quantum field theory}

In recent times, the kinematic quantitites (expansion, shear and rotation),
as well as the Raychaudhuri equations, have appeared, quite unexpectedly,
in the context of quantum field theory. We outline below, some of these
scenarios briefly.

\subsubsection{The Langevin--Raychaudhuri equation}

A couple of years ago, Borgman and Ford investigated  gravitational effects of 
quantum stress tensor fluctuations 
\cite{ford2004}, \cite{borgman}.
They showed that these fluctuations produce fluctuations in the focusing of a 
bundle of geodesics. An explicit calculation using the Raychaudhuri 
equation, treated as a Langevin equation (ignoring the $\Theta^2$ term by a
smallness assumption converts the Raychaudhuri equation to the form
$\frac{d\Theta}{d\lambda} = f(\lambda)$, a Langevin--type equation) 
was performed to estimate
angular blurring and luminosity fluctuations of the images of 
distant sources. The stress--tensor fluctutations were obtained assuming the 
case of a massless, minimally coupled scalar field 
in a flat background in a thermal state. Scalar field fluctuations drove the
Ricci tensor fluctuations (via the semiclassical Einstein equations), which,
in turn led to fluctuations in the expansion $\Theta$.
These authors also made some numerical estimates for the quantity 
$\frac{\Delta L}
{L}$ (the fractional luminosity fluctuation) and pointed out possible 
astrophysical situations (gamma--ray bursts, for example), where 
an observable value of this quantity might exist. However, it is true that
their work has many assumptions (ignoring shear, rotation as well as the
$\Theta^2$ term etc.) and much further analysis is required to have a
clearer picture of the possibility iof actually seeing such effects in
the real universe. Nevertheless, the idea, on the whole, is novel and
interesting in its own right and provides us with another application of
the Raychaudhuri equation in a very different scenario.

\subsubsection{RG flows in theory (coupling) space}

Till now we have been looking at geodesic flows in spacetime. However,
one might also contemplate geodesic flows in fictitious spaces where
a metric is defined. Such spaces may correspond to certain physical
scenarios. We highlight one example here in order to show how
ubiquitous the notions of expansion, rotation, shear and the Raychaudhuri
equation are.

For a given Lagrangian field theory, the set of couplings can form
a space which is known as {\em theory/coupling space}. A curve in such
a space will therefore represent a flow of couplings. In such a space,
we can define a distance function using two--point functions
integrated over physical spacetime. Such a definition of metric
goes back to the well-known Cramer--Rao metric in probability theory
and has also been highlighted in the context of field theory by
Zamolodchikov as well as O'Connor--Stephens (see \cite{skar01} 
and references therein). 

Once we have a metric, we can use it to define derivatives of a vector
field. The vector field of interest here is the so--called $\beta-function$
vector field, which generates the RG flow. The covariant derivative of this 
vector field when split into the usual trace, anti--symmetric and symmetric
traceless parts define the usual expansion, rotation and shear. However, we 
also know that
the $\beta$--function vector field must satisfy the conformal Killing
condition (this is the geoemtric form of the RG equation). These facts together imply the result that the focal point
of the congruence generated by the $\beta$-function vector field 
must necessarily be a fixed point (i.e. all $\beta_a$ (`a' labels the
coupling space coordinates) vanish there) \cite{skar01}

This line of thought is useful in obtaining generic results in field theory
--the understanding of shear and rotation in the context of RG flows
might provide a better geometric understanding of RG flows. 

\subsubsection{Holography, c--function and the Raychaudhuri equation for the 
expansion}

The holographic principle has recently played a crucial role in our 
understanding of quantum aspects of gravity. The principle states that the 
information of gravity 
degrees of freedom in a D dimensional volume is encoded in a quantum field 
theory defined on the $(D-1)$ dimensional boundary of this volume. 
As an immediate 
realization of this, it is shown that the Renormalization Group flow equation 
for the $\beta$--function of a four dimensional 
quantum gauge field theory defined on the boundary of a five dimensional volume 
can be described by geodesic congruences in a scalar--coupled five
dimensional gravitational theory. Such a gauge-gravity duality was proposed 
in a more general framework through the Maldacena conjecture and can be 
elegantly described 
through the Renormalization Group equation with the bulk coordinate 
(or the holographic coordinate)
as the Renormalization 
Group parameter. It is shown that if the central charge or the c-function in
a quantum field theory evolves monotonically under RG then the holographic 
principle 
indicates that in the corresponding dual gravity in five dimensions,
the picture is realized through the
Raychaudhuri equation governing the monotonic flow of the expansion 
parameter $\Theta$ for the geodesic congruences in the gravity sector
\cite{bgm2000}, \cite{sahakian2000}\cite{alvarez1999}. 
The central charge of the 
boundary theory, or the c-function, is  a measure of the degrees of freedom 
of the theory. As a consequence 
it is also a measure of the entropy of the black hole in the dual gravity 
sector which is 
related to the area of the horizon through the Bekenstein-Hawking formula. 
The effective central charge 
is a function of the couplings of the theory, which monotonically decreases 
as one flows to lower energies 
through the RG equation. The fixed point described by the 
boundary conformal field theory corresponds 
to the extrema of this function. It turns out that the null geodesic 
congruences can be used 
as a probe to decode the holographically encoded informations.

Consider a D-dimensional spacetime with a negative cosmological constant 
where the metric $g_{ab}$ is 
foliated by appropriate choices of constant time surfaces. If we now 
focus on a $(D-2)$ dimensional 
spacetime surface $M$ at a fixed time t, then one may construct a null vector 
field $m^a$ on the 
light sheets (consisting of spacetime points scanned by null geodesic 
congruences) which is orthogonal to M 
such that $m^an_a = -1$, where $n_a$ are tangents to the geodesic. 
The metric on M is given by
\begin{equation}
h_{ab} = g_{ab} + n_{(a}m_{b)}
\end{equation}
Defining,
\begin{equation}
B_{ab} = D_{b} n_{a}
\end{equation}
such that $B_{ab} = B_{ba}$ (Frobenius theorem ensures the symmetry property) 
and 
\begin{equation}
\Theta = Tr B
\end{equation}
it can be shown from the Raychaudhuri equation for the expansion that 
the geodesics converge since,
\begin{equation}
\Theta \leq 0
\end{equation}
The question that arises now is -- how is the information on the light 
sheet encoded in M?
Assuming that the spacetime admits  a timelike Killing vector field, we may 
visualize the 
time flow of M to constitute the $(D-2)$ + 1 = $(D-1)$ dimensional boundary 
of the $D$ dimensional bulk. 
The  c--function of the corresponding dual gauge theory sits on the boundary. 
The RG flow to 
lower energy scales is then associated to the motion along the converging 
congruences of null geodesics 
described by Raychaudhuri equation. In this sense, the null geodesics act as 
a probe for the dual theory at a lower 
energy scale. Pictorially, it may be described as follows. The lower the 
energy scale, the RG takes us into the gauge theory,
deeper inside the bulk we move along null geodesic congruences following 
Raychaudhuri equation. The caustic 
point where the geodesics meet correspond to the fixed point of the RG flow.
This establishes a remarkable significance of the Raychaudhuri equation in 
describing the holographic principle.

As described previously the c-function tells us  about the degree of freedom 
of the system, which, in turn, is 
related to black hole entropy through the area law. Therefore, the c-function 
is naturally related to the area function 
$A(r)$ as,
\begin{equation}
c(r) =  A(r)/4
\end{equation}
In the context of the recently developed  attractor mechanism \cite{cvetic}in describing black holes in a scalar coupled gravitational theory, 
one uses this flow of the c-function to decode some interesting properties of
both supersymmetric and non-supersymmetric attractors. The  
Raychaudhuri equation once 
again plays a crucial role.It is shown that in any spherically symmetric scalar coupled static asymptotically flat
solution, $c(r)$ decreases monotonically as one moves radially inward from infinity. It is further shown that the 
minimum value of $c(r)$ corresponds to the entropy of the horizon\cite{sandip}.
In order to establish the c-theorem in this context
once again the Raychaudhuri equation is used. Taking a congruence of null 
geodesics, the expansion is
given by,
\begin{equation}
\Theta = \frac{dlnA}{d\lambda}
\end{equation}
(where $\lambda$ is the affine parameter). Using the null energy condition,
\begin{equation}
T_{ab}k^a k^b \geq 0
\end{equation}
and the Raychaudhuri equation, we get
\begin{equation}
\frac{d\Theta}{d\lambda} \leq 0
\end{equation}
The c-theorem can thus be proved on very general grounds. 

In summary, these rather novel applications of the equations in the
context of quantum field theory seems to assert once again the 
generality and the ubiquitous nature of the equations.

\section{Summary and scope}

We have reviewed, in this article, the Raychaudhuri equations and its
applications in diverse contexts. It must be admitted that since its
inception, the equations have used extensively to enhance our understanding
of various situations within, as well as outside the scope of GR. 
The primary reason behind its wide--ranging applicability is (as
emphasised several times in this article) the fact that the equations
encode {\em geometric} statements about flows. Since flows appear
in many different contexts in physics, it goes without saying, that
the equations will be useful in furthering our understanding in
different ways. 

Among the applications within classical GR, the utility of the
equations in understanding the singularity 
problem, is surely the most prominent one. In astrophysics, we have 
quoted the use of the equations in the context of lensing, cracking of
self--gravitating objects. At a more fundamental level, Jacobson's work
exploits the equation and, in some sense, makes use of its {\em geometric}
nature to arrive at a new way of looking at the Einstein equations. 
 
On the other hand, in a non--Riemannian spacetime, we have seen how the 
equations change
--in particular, in spacetimes with torsion. Torsion appears in disguise
in string theory and is therefore not totally irrelevant today, even though
observations might have ruled out the Einstein--Cartan--Sciama--Kibble theory
long ago. 

Furthermore, what has come as a pleasant surprise in recent times, is the 
utility of these
equations (and the quantities that appear in the equations: expansion,
shear and rotation) in the context of quantum field theory. In addition,
the generalisation of these equations to surface congruences is also
an interesting development, despite the fact that not much has been
done on such generalised Raychaudhuri (Capovilla--Guven) equations.

Despite the large body of work on or based on the Raychaudhuri equations, 
there still remain many un--answered questions. 
In conclusion, we list a few of them.
The choice is surely biased by our own perspective.

(a) Normally, in fluid mechanics, the independent variable is the velocity field.
The Raychaudhuri equations are for the gradient of the velocity field--they
are one derivative higher.
We mentioned before that they are essentially identities and become 
equations when we choose a geometric property (eg. vacuum, Einstein space etc.) or use Einstein's field equations of GR. Is it possible to set-up and solve 
the initial value problem for this coupled system of equations such that
we know the characteristics of a geodesic flow in a given spacetime geometry
once we specify the initial conditions on expansion, shear and rotation?
Similar analysis is done while deriving the focusing theorem but can we
do it by analysing the full system of equations analytically/numerically.  
Following this approach one might be able to reconstruct the gradient of the 
velocity field and also obtain the velocity field itself.

(b) In the context of RG flows, is it possible to understand the role of
shear and rotation and use it to improve our geometric analysis of such 
flows?

(c) For surface congruences, can we obtain equations for families of null
surfaces? Or, for surfaces defined by the extremality of actions other
than Nambu--Goto (rigidity corrections, Willmore functionals etc.)

(d) In quantum mechanics, we talk about the flow of probability. Imagining this
as a fluid, can we obtain expansion, rotation and shear of such flows and
then write down conclusions based on the corresponding Raychaudhuri equations?

(e) Electrodynamics, is, after all, the physics of the electric and magnetic
vector fields. Suppose we construct the gradient of the electric/magnetic
fields, split it up into expansion, rotation and shear and write down
Raychaudhuri equations for the so--called Faraday lines of force! What does
such an analysis tell us? 

To sum up, we would like to {\em conjecture} that wherever there are vector 
fields
describing a physical/geometrical quantity, there must be corresponding 
Raychaudhuri equations. We believe, there are 
many {\em avatars} of the same equations, of which, till today, we are aware 
of only a few.

\section*{Acknowledgements}

The authors thank the editors (N. Dadhich, P. S. Joshi and Probir Roy) 
of this special volume, for giving them the
opportunity to write this review. They also specially thank Probir Roy and
Anirvan Dasgupta for 
their comments and a careful reading of the manuscript.

%\begin{figure}
%\centerline{\epsfxsize=6in\epsffile{revsum.eps}}
%\caption{Summary}
%\end{figure}


\begin{references}

\bibitem{akr1953} A. Raychaudhuri, Phys. Rev. {\bf 89}, 417
(1953) 
\bibitem{akr1953a} A. Raychaudhuri, Phys. Rev. {\bf 86}, 90
(1952) 
\bibitem{akr1955} A. Raychaudhuri, Phys.
Rev. {\bf 98}, 1123 (1955)
\bibitem{akr1957} A. Raychaudhuri, Z. Astrophysik, {\bf 43}, 161 (1957) 
\bibitem{hs1955} O. Heckmann and E. Schucking, {\em } Z. Astrophysik
{\bf 38}, 95 (1955)
\bibitem{komar1956} A. Komar, {\em } Phys. Rev. {\bf 104}, 544 (1956)
\bibitem{akr1957b} A. Raychaudhuri, Phys. Rev. {\bf 106}, 172 (1957) 
\bibitem{ehlers} J. Ehlers, {\em Contributions to the relativistic
mechanics of continuous media}, Gen. Rel. Grav. {\bf 25}, 1225 (1993)
, {English translation of original German article by P. Jordan, J. Ehlers,
W. Kundt, R. K. Sachs, Proceedings of the Mathematical--Natural
Sciences Section of the Mainz Academy of Sciences and Literature, Nr. 11, 
792 (1961)} 
\bibitem{sachs} R. K. Sachs, Proc. Roy. Soc. (London) {\bf A 264}, 309
(1961); {\em ibid} {\bf A 270}, 103 (1962)
\bibitem{ll} L. Landau and E. M. Lifshitz, {\em Classical theory of
fields}, (Pergamom Press, Oxford, UK, 1975)
\bibitem{lk1963} E. M. Lifshitz and I. M. Khalatnikov, Adv. Phys. {\bf 12}, 185 (1963)
\bibitem{rp1965}  R. Penrose, Phys. Rev. Lett. {\bf 14}, 57 (1965) 
\bibitem{swh1965} S. W. Hawking, Phys. Rev. Lett. {\bf 15}, 689 (1965)
\bibitem{swh1966} S. W. Hawking, Phys. Rev. Lett. {\bf 17}, 444 (1966)
\bibitem{wald} R. M. Wald, {\em General Relativity}, (University of Chicago
Press, Chicago, USA, 1984)
\bibitem{he1973} S. W. Hawking and G. F. R. Ellis, {\em The large scale
structure of spacetime}, (Cambridge University Press, Cambridge, UK, 1973)
\bibitem{joshi} P. S. Joshi, {\em Global aspects in gravitation and cosmology}, (Oxford University Press, Oxford, UK, 1997)
\bibitem{poisson} E. Poisson, {\em A relativist's toolkit: the mathematics of
black hole mechanics}, (Cambridge
University Press, Cambridge, UK, 2004)
\bibitem{ellis1} G. F. R. Ellis in {\em General Relativity and Cosmology},
{\em International School of Physics, Enrico Fermi--Course XLVII}, (Academic
Press, New York, 1971)
\bibitem{ciufo} I. Ciufolini and J. A. Wheeler, {\em Gravitation and
Inertia}, (Princeton University Press, Princeton, USA, 1995)
\bibitem{wormholes} M. Visser, {\em Lorentzian wormholes from Einstein to
Hawking}, (AIP Press, USA, 1995)
\bibitem{identity} T. Frankel, {\em Gravitational curvature}, (W. H. Freeman,
USA, 1979)
\bibitem{zafiris} E. Zafiris, J.Geom.Phys. {\bf 28}, 271 (1998) 
\bibitem{carter} B. Carter, Contemp.Math. {\bf 203}, 207 (1997)
\bibitem{fjt1978} F. J. Tipler, Phys. Rev. {\bf D17}, 2521 (1978)
\bibitem{fjt1978a} F. J. Tipler, J. Diff. Equ. {\bf 30}, 165 (1978)
\bibitem{jdb1976} J. D. Bekenstein, Phys. Rev. {\bf D 11}, 2072 (1975)
\bibitem{glm1973} G. L. Murphy, Phys. Rev. {\bf D 8}, 4231 (1973)
\bibitem{egy} H. Epstein, V. Glaser and A. Jaffe, Nuovo Cimento {\bf 36},
1016 (1965)
\bibitem{darkenergy} R. R. Caldwell, M. Kamionkowski and N. N. Weinberg,
Phys. Rev. Lett. {\bf 91}, 071301 (2003);
S. M. Carroll, M Hoffman and M. Trodden, Phys. Rev. {\bf D68}, 023509 (2003);
P. Frampton, Phys. Lett. {\bf B555}, 139 (2003);
V. Sahni and Y. Shtanov, JCAP {\bf 0311}, 014 (2003)
\bibitem{rs} L. Randall and R. Sundrum, Phys. Rev. Lett. {\bf 83}, 4690 (1999);
C. Csaki, {\em TASI Lectures on Extra Dimensions and Branes} hep-ph/0404096 
\bibitem{scg} N. D. Birrell and P. C. W. Davies, {\em 
Quantum fields in curved space}, (Cambridge University Press, Cambridge, UK, 1982) 
\bibitem{casimir} H. B. G. Casimir, {\em On the attraction between two
perfectly conducting plates}, Proc. Kon. Nederl. Akad. Wetenschap, {\bf 51},
793 (1948)
\bibitem{ab1987} A. Borde, Class. Qtm. Grav. {\bf 4}, 343 (1987)
\bibitem{tr1988} T. Roman, Phys. Rev. {\bf  D 37}, 546 (1988)
\bibitem{fjt1977} F. J. Tipler, Ann. Phys. {\bf 108}, 1 (1977); {\em ibid.} Phys. Rev. Lett. {\bf 37}, 879 (1976)
\bibitem{ce1980} C. Chicone and P. Ehrlich, Manuscripta Math. {\bf 31}, 297 (1980)
\bibitem{rb1983} T. Roman and P. G. Bergmann, Phys. Rev. {\bf D28}, 1265 (1983)

\bibitem{tr1986} T. Roman, Phys. Rev. {\bf D 33}, 3526 (1986)
\bibitem{viss1992} M. Visser, Phys. Rev. {\bf D47}, 2395 (1993)
\bibitem{akr1998} A. K. Raychaudhuri, Phys. Rev. Lett. {\bf 80}, 654 (1998);
A. Saa, Phys. Rev. Lett. {\bf 81}, 5031 (1998); J. M. M. Senovilla, Phys. Rev. Lett. {\bf 81}, 5032 (1998); A. K. Raychaudhuri, Phys. Rev. Lett. {\bf 81}, 
5033 (1998)
\bibitem{akr2000} A.K. Raychaudhuri, Mod. Phys. Lett. {\bf A15}, 391 (2000) 
\bibitem{senovilla} J. M. M. Senovilla, Phys. Rev. Lett. {\bf 64}, 2219
(1990); E. Ruiz and J. M. M. Senovilla, Phys. Rev. {\bf D45}, 1995 (1992)
\bibitem{jose} J. M. M. Senovilla, article in this volume.
\bibitem{akr1974} A. K. Raychaudhuri and S. Banerji, Zeit. fur. Astrophysik
{\bf 58}, 187 (1964)
\bibitem{sb1974} S. Banerji, Phys. Rev. {\bf D 9}, 877 (1974)
\bibitem{akr1975} A. K. Raychaudhuri, Prog. Theor. Phys. {\bf 53}, 1360 (1975)
\bibitem{kung1} J. H. Kung, Phys. Rev.{\bf D 52}, 6922 (1995)
\bibitem{muller} V. Muller and H. J. Schmidt, Gen. Rel. Grav. {\bf 17}, 769
(1985); D. Page, Phys. Rev. {\bf D 36}, 1607 (1987); J. D. Barrow and A. Ottewill, J. Phys. A: Math. Gen. {\bf 16}, 2757 (1983)
\bibitem{kung} J. H. Kung, Phys.Rev. {\bf D 53}, 3017 (1996)
\bibitem{whitt} B. Whitt, Phys. Lett. {\bf 145B}, 176 (1984); J. C. Alonso, F. Barbero, J. Julve, and A. Tiemblo, Class. Qtm.
Grav. {\bf 11}, 865 (1994); M. Mijic et al., Phys. Rev. {\bf D 34}, 2934 (1986).

\bibitem{schouten} J.A. Schouten {\it Ricci Calculus}
(Springer-Verlag, Berlin, 1954)
\bibitem{hehl}
F.W. Hehl, P. von der Heyde, G.D. Kerlick and J.M. Nester, Rev. Mod. Phys. 
{\bf 48}, 393 (1976). 
\bibitem{GSW} M.B. Green, J.H. Schwarz, E. Witten, 
{\it  Superstring Theory}, (Cambridge University Press, Cambridge, UK, 1987);
R. Hammond,  Il Nuovo Cimento {\bf 109B}, 319 (1994);
          Gen. Rel. Grav. {\bf 28}, 419 (1986);
 V. De Sabbata,  Ann. der Physik {\bf 7}, 419 (1991);
 V. de Sabbata,  {\it Torsion, string tension and quantum gravity},
           in ``Erice 1992'', Proceedings, String Quantum Gravity and Physics
           at the Planck Energy Scale, p. 528;
 Y. Murase,  Prog. Theor. Phys. {\bf 89}, 1331 (1993)
\bibitem{HOW}
P.S. Howe, A. Opfermann, G. Papadopoulos, {\it Twistor
spaces for QKT manifold}, hep-th/9710072; P.S. Howe, G.
Papadopoulos, Phys. Lett. {\bf 379B}, 80 (1996); V. de Sabbata, C.
Sivaram, {\it Il Nuovo Cimento} {\bf 109A}, 377 (1996)
\bibitem{LOS}
G. Papadopoulos, P.K. Townsend, {\it Nucl. Phys.} {\bf B444}, 245 (1995);
C.M. Hull, G. Papadopoulos, P.K. Townsend, Phys. Lett. {\bf 316B} 291 (1993);
B.D.B. Figueiredo, I. Damiao Soares and J. Tiomno, Class. Qtm. Grav.
{\bf 9}, 1593 (1992); L.C. Garcia de Andrade, Mod. Phys.
Lett. {\bf 12A} 2005 (1997); O. Chandia, Phys. Rev. {\bf D55}, 7580 (1997); P.S. Letelier, Class. Quant. Grav. {\bf 12}, 471 (1995); 
P. S. Letelier, Class. Qtm. Grav. {\bf 12}, 2221 (1995); 
R. W. K\"uhne, Mod. Phys. Lett. {\bf A12}, 2473 (1997); 
D. K. Ross, Int. J. Theor. Phys. {\bf 28}, (1989) 1333.  
\bibitem{we} S. Kar, S. SenGupta, S. Sur, Phys. Rev. {\bf D67}, 044005 (2003); 
S. Kar , P. Majumdar,
S. SenGupta, S. Sur, Class. Qtm. Grav. {\bf 19}, 677 (2002); 
S. SenGupta, S. Sur, Phys. Lett. {\bf B521}, 350 (2001);
S. Kar , P. Majumdar, S. SenGupta , A. Sinha, Eur. Phys. Jr. {\bf C23}, 357 (2002); 
P. Majumdar, S. SenGupta, Class. Qtm. Grav.{\bf 16}, L89 (1999);
S. SenGupta , S. Sur, Europhys. Lett. {\bf 65}, 601 (2004)
\bibitem{capoz} S. Capozziello, G. Lambiase , C. Stornaiolo, Ann. der Phys. {\bf 10}, 713 (2001), gr-qc/0101038
\bibitem{drls} R. de Ritis, M. Lavorgna, C. Stornaiolo, Phys.Lett. {\bf
95A}, 425 (1983)
\bibitem{trautman}
A. Trautman, Nature (Phys. Sci.)  {\bf 242}, 7 (1973) 
\bibitem{stewart}
J. Stewart and P. Hajicek, Nature (Phys. Sci.) {\bf 244}, 96 (1973)
\bibitem{esposito} G. Esposito, Fortschritte der Physik {\bf
40}, 1 (1992)
\bibitem{ddrpss} M. Demianski, R. de Ritis, G. Platania, P. Scudellaro and C. Stornaiolo, Phys. Rev. {\bf D35} 1181 (1987)
\bibitem{Ellis} G. F. R. Ellis and M. Bruni, Phys.  Rev. {\bf D40}, 1804 (1989);  G. F. R. Ellis, J. Hwang and M. Bruni, Phys.  Rev. {\bf
D40}, 1819 (1989); G. F. R. Ellis, M. Bruni and J. Hwang, Phys. Rev. {\bf D42}, 1035 (1990)
\bibitem{palle} D. Palle, Nuovo Cim. {\bf B114}, 853 (1999)
\bibitem{tafel2} J. Tafel, Phys. Lett. {\bf A45}, 341 (1973)
\bibitem{raych} A.K. Raychaudhuri, {\em  Theoretical
Cosmology}, (Clarendon Press, Oxford, UK, 1979)

\bibitem{void2000} N. Sugiura, K. Nakao, D. Ida, N. Sakai, H. Ishihara, 
Prog. Theor. Phys. {\bf 103}, 73 (2000) 

\bibitem{herrera} L. Herrera, Phys. Letts. {\bf A165}, 206 (1992)

\bibitem{prisco} A. di Prisco, E. Fuenmayor, L  Herrera, V Varela,
Phys. Lett. {\bf A 195}, 23 (1994)

\bibitem{jacobson1} T. Jacobson, Phys. Rev. Lett. {\bf 75}, 1260 (1995)
\bibitem{jacobson2} C. Eling, R. Guedens, and T. Jacobson, Phys. Rev. Lett. {\bf 96}, 121301 (2006)

\bibitem{lyth} D. Lyth, M. Mukherjee, Phys. Rev. {\bf D38}, 485 (1988)

\bibitem{nouri2006} N. Ahmadi, M. Nouri-Zonoz, {\em 
Quantum gravitational optics: Effective Raychaudhuri equation},
gr-qc/0605009

\bibitem{tsagas} C. Tsagas, {\em Magnetic tension and gravitational
collapse}, gr-qc/0311085 

\bibitem{maartens} R. Maartens, 
Phys. Rev. {\bf D 62}, 084023 (2000)

\bibitem{cg1995} R. Capovilla and J. Guven,
Phys. Rev. {\bf D52}, 1072 (1995); S. Kar, Phys. Rev. {\bf D52}, 2036 (1995);
S. Kar, Phys. Rev. {\bf D53}, 2071 (1996)

\bibitem{sayan1} S. Kar, Phys. Rev. {\bf D55}, 7921 (1997) 

\bibitem{sayan2} S. Kar, {\em Focusing of branes in warped backgrounds}, Ind. J. Phys. (to appear in the special issue commemorating A. K. Raychaudhuri) (2006)

\bibitem{sayan3} S. Kar, Phys.Rev. {\bf D54}, 6408 (1996) 

\bibitem{sayan4}
S. Kar, {\em  Generalised Raychaudhuri Equations for Strings and Membranes}
Proceedings of IAGRG--XVIII (Matscience, Madras, India (Feb. 15-17, 1996),
Matscience Report no. 117)
     
\bibitem{ford2004}
J. Borgman and L. H. Ford, Phys.Rev. {\bf D70}, 064032 (2004) 

\bibitem{borgman} J. Borgman, {\em Fluctuations of the expansion: 
The Langevin-Raychaudhuri equation}, Ph.d. thesis, Tufts University, USA (2004)

\bibitem{skar01} S. Kar, Phys. Rev. {\bf D64}, 105017 (2001) 

\bibitem{bgm2000} V. Balasubramanian, E. Gimon, D. Minic, JHEP {\bf 05}, 014 (2000)

\bibitem{sahakian2000} V. Sahakian, Phys.Rev. {\bf D62}, (2000) 126011 

\bibitem{alvarez1999} E. Alvarez, C. Gomez,  Nucl.Phys. {\bf B541}, 441 
(1999) {\em ibid.} {\em Holography and the c-theorem} hep-th/0009203

\bibitem{cvetic} M. Cvetic and A. A. Tseytlin, Phys. Rev. {\bf D 53}, 5619 (1996); A.Stromonger, Phys. Lett. {\bf B 383}, 1996; S. Ferrara and R. Kallosh, 
Phys. Rev. {\bf D 54}, 1525 (1996)
S. Ferrara, G. W. Gibbons and R. Kallosh, Nucl. Phys. {\bf B 500}, 75 (1997). 
\bibitem{sandip} K. Goldstein, R. P. Jena, G. Mandal and
S. P. Trivedi, JHEP {\bf 02}, 053 (2006)
\end{references}
\end{document}